\documentclass[aps,prl,twocolumn,superscriptaddress]{revtex4}
\usepackage{graphicx}
\usepackage{latexsym}
\usepackage{amssymb}
\usepackage{amsmath}
\usepackage{amsfonts}
\usepackage{bm}
\usepackage{multirow}
\usepackage{color}
\newcommand{\ii}{\mathrm{i}}
\newcommand{\dd}{\mathrm{d}}

\newcommand{\dsZ}{\mathbb{Z}}

\newcommand{\Tr}{\mathop{\mathrm{Tr}}}
\renewcommand{\Re}{\mathop{\mathrm{Re}}}

\newcommand{\vect}[1]{{\bm{#1}}}
\newcommand{\eqnref}[1]{Eq.\,\eqref{#1}}
\newcommand{\figref}[1]{Fig.\,\ref{#1}}
\newcommand{\tabref}[1]{Tab.\,\ref{#1}}

\newcommand{\beq}{\begin{equation}}
\newcommand{\eeq}{\end{equation}}
\newcommand{\beqn}{\begin{eqnarray}}
\newcommand{\eeqn}{\end{eqnarray}}

\begin{document}

\title{Topological Orders with Global Gauge Anomalies}



\author{Yi-Zhuang You}

\author{Cenke Xu}


\affiliation{Department of physics, University of California,
Santa Barbara, CA 93106, USA}

\begin{abstract}

By definition, the physics of the $d-$dimensional (dim) boundary
of a $(d+1)-$dim symmetry protected topological (SPT) state cannot
be realized as itself on a $d-$dim lattice. If the symmetry of the
system is unitary, then a formal way to determine whether a
$d-$dim theory must be a boundary or not, is to couple this theory
to a gauge field (or to ``gauge" its symmetry), and check if there
is a gauge anomaly. In this paper we discuss the following
question: can the boundary of a SPT state be driven into a fully
gapped topological order which preserves all the symmetries? We
argue that if the gauge anomaly of the boundary is
``perturbative", then the boundary must remain gapless; while if
the boundary only has global gauge anomaly but no perturbative
anomaly, then it is possible to gap out the boundary by driving it
into a topological state, when $d \geq 2$. We will demonstrate
this conclusion with two examples: (1) the $3d$ spin-1/2 chiral
fermion with the well-known Witten's global
anomaly~\cite{wittensu2}, which is the boundary of a $4d$
topological superconductor with SU(2) or U(1)$\rtimes Z_2$
symmetry; and (2) the $4d$ boundary of a $5d$ topological
superconductor with the same symmetry. We show that these boundary
systems can be driven into a fully gapped $\mathbb{Z}_{2N}$
topological order with topological degeneracy, but this
$\mathbb{Z}_{2N}$ topological order cannot be future driven into a
trivial confined phase that preserves all the symmetries due to
some special properties of its topological defects.

\end{abstract}


\maketitle

\section{1. Introduction}

The contrast between bulk and boundary is the most general and
important feature of all symmetry protected topological (SPT)
states. A SPT state has a gapped and nondegenerate bulk state, but
it must also have a nontrivial boundary. A nontrivial boundary
must satisfy two criteria: (1) the boundary must be either gapless
or degenerate, as long as the symmetry of the system is not
explicitly broken; (2) the low energy physics of the boundary
cannot be realized as a lower dimensional system as itself. For
example, in the noninteracting case, the $2d$ boundary of the $3d$
topological insulator is a single (or odd number of) massless $2d$
Dirac fermion, which cannot exist in any $2d$ free fermion lattice
model with time-reversal and charge U(1) symmetry, and it will
remain gapless as long as both symmetries are
preserved~\cite{fukane,moorebalents2007,roy2007}.

The second criterion of SPT states is especially important, it
implies that if we attempt to regularize the boundary of a SPT
state as a lower dimensional system, some ``anomaly" will occur.
The most well-known example of anomaly is the U(1) gauge anomaly
of chiral fermions in odd spatial dimensions. For example, let us
consider a $1d$ left-moving complex chiral fermion, and let us
assume there is an exact U(1) symmetry associated with the charge
conservation (this exact U(1) symmetry is an important assumption
in the no-go theorem proved in Ref.\,\cite{doublingA,doublingB}).
If this U(1) symmetry exists in the fully regularized lattice
model, then there should be no problem of enhancing this global
U(1) symmetry to a local U(1) gauge symmetry, $i.e.$ we should be
able to couple this chiral fermion to a U(1) gauge field. However,
it is well-known that a chiral fermion coupled to U(1) gauge field
will have gauge anomaly: namely the gauge current is no longer
conserved: $\partial_\mu j_\mu \sim F_{01}$, which causes
inconsistency (anomaly). This anomaly implies that a $1d$ chiral
fermion can only exist at the boundary of a $2d$ system, and the
physical interpretation of the chiral anomaly is merely the
quantum Hall physics: charge is accumulated at the boundary when
magnetic flux is adiabatically inserted in the $2d$ bulk. The
anomaly of the boundary of $3d$ topological insulator was
discussed in Ref.~\onlinecite{maxfisher}. More general relation
between boundary anomaly and bulk SPT states has been studied
systematically in Ref.~\onlinecite{wenanomaly}.

Generally speaking, bulk states and boundary states do not have
one-to-one correspondence, $i.e.$ the bulk state does not uniquely
determine its boundary, but the boundary state will determine the
bulk state~\cite{kongwen}. The boundary state of a SPT state
depends on the Hamiltonian at the boundary, or in other words
depends on how the bulk Hamiltonian ``terminates" at the boundary.
Thus different boundary states can belong to the same
``universality class", if they correspond to the same bulk state.
Different boundary states belonging to the same universality class
must share the same universal properties, and these universal
properties are precisely the ``anomalies".

Based on the definition of SPT states, the boundary of all $1d$
SPT states must be degenerate; the boundary of all $2d$ SPT states
must be either gapless or spontaneously break certain discrete
symmetry which leads to ground state degeneracy; the boundary of
SPT states on three and higher spatial dimensions has even richer
possibilities: besides gapless spectrum and spontaneous symmetry
breaking, the boundary can also have fully gapped topological
order which preserves all the symmetries of the system. This last
possibility is what we will study in this paper. The boundary
topological order, although gapped, must still be anomalous,
namely it cannot be realized as a lower dimensional system itself.
For example, the ``anomalous" boundary topological order of $3d$
topological insulator and topological superconductor $^3$He-B
phase has already been
studied~\cite{TI_fidkowski1,TI_fidkowski2,TI_qi,TI_senthil,TI_max}.
And one natural boundary state of a $3d$ bosonic SPT
state~\cite{wenspt,wenspt2} is a $Z_2$ topological orders whose
$e$ and $m$ excitations carry fractional quantum numbers of the
symmetry~\cite{senthilashvin,xuclass} (some systems can also have
a different boundary topological order with
semions~\cite{fiona2014}), and this particular kind of
fractionalization cannot exist in $2d$, even though it is
consistent with all the fusion rules of $e$ and $m$ excitations.

In this paper we will focus on the SPT states whose symmetry group
$G$ is unitary. With unitary symmetries, there is a formal way to
determine if a $d-$dimensional low energy theory is anomalous or
not: we can couple the system to gauge field with gauge group $G$
(or in other words we ``gauge" the symmetry $G$), and check if
there is any gauge anomaly. This procedure does not directly apply
to the $2d$ boundary of many $3d$ TI/TSC, because these systems
usually involve a nonunitary time-reversal symmetry which cannot
be ``gauged".

Gauge anomaly is very well studied in high energy physics. It
turns out that there is a {\it precise} correspondence between the
free fermion topological insulator/superconductor with unitary
symmetry $G$ in $(d+1)-$dimensional space, and the gauge anomaly
at its $d-$dimensional boundary space after gauging: if the bulk
classification is $\mathbb{Z}$, its boundary must have
perturbative gauge anomaly; if the bulk classification is
$\mathbb{Z}_2$, its boundary must have global gauge anomaly after
gauging. Notice that for TSC with no symmetry at all, because its
boundary modes can only couple to gravitational field, its
boundary has a precise correspondence with gravitational anomalies
computed in Ref.~\onlinecite{wittenanomaly}, see \tabref{tab:
anomaly}.

Not all SPT states can have fully gapped symmetric boundary
topological order, even in dimensions higher than $3$. First of
all, if after gauging, the boundary of a SPT state has
perturbative anomaly (such as U(1) chiral anomaly) which can be
calculated using standard perturbation theory, it can never be
gapped out into a boundary topological order, because this
boundary must respond to weak background gauge field
configurations, thus the boundary must remain gapless as long as
its symmetry is preserved. The well-known ``anomaly matching
condition" was meant to deal with the perturbative anomaly
only~\cite{anomalymatching,anomalymatching2}, although the concept
of topological order was not developed by then. This conclusion
will be further demonstrated with concrete examples in the next
section. However, if after gauging a boundary has global gauge
anomaly, then gaplessness is no longer a necessity, which means
that it is at least possible to drive the boundary into a fully
gapped topological order which inherits the global anomaly. In
this paper we will investigate the only two systems in
\tabref{tab: anomaly} with global gauge anomalies: the Witten's
anomaly~\cite{wittensu2} in $(3+1)d$ (boundary of a $(4+1)d$
topological superconductor), and the analogue of Witten's anomay
in $(4+1)d$ (boundary of a $(5+1)d$ system). We will demonstrate
that it is possible to drive these boundary systems into a
topological order, and the topological order cannot be further
driven into a gapped nondegenerate symmetric trivial confined
phase, because of their anomalies.

We also note that in Ref.~\onlinecite{senthilhe3}, a ``symmetry
enforced gapless" state is proposed for the $2d$ boundary of the
$3d$ fermionic SPT state with SU(2) and time-reversal symmetry.
The authors argued that this boundary cannot be gapped into a
topological order with the full SU(2) and time-reversal symmetry.
We want to stress that in our paper we restrict our discussion to
the cases with unitary symmetries, so that we can ``gauge" all the
symmetries, and make a precise comparison between the
classification of TI/TSC with the well-known gauge anomalies.

\begin{table}[htdp]
\caption{The correspondence between bulk classifications of
noninteracting topological insulator (TI) and topological
superconductor (TSC) in each spatial dimension
$d$~\cite{ludwigclass1,ludwigclass2,kitaevclass} and gauge
anomalies at $d-1$ dimensional boundary. This table is periodic
with periodicity 8. The first row corresponds to the TSC without
any symmetry, thus its boundary can only have gravitational
anomaly. G and P stand for global and perturbative anomalies
respectively.}
\begin{center}
\begin{tabular}{c p{17pt} p{17pt} p{17pt} p{17pt} p{17pt} p{17pt} p{17pt} p{17pt} p{17pt} p{17pt}}
\hline
\multicolumn{11}{c}{Bulk TI/TSC classification}\\
$d$ & 1 & 2 & 3 & 4 & 5 & 6 & 7 & 8 & 9 & 10 \\
\hline
none \ \  & $\dsZ_2$ & $\dsZ$ & 0 & 0 & 0 & $\dsZ$ & 0 & $\dsZ_2$ & $\dsZ_2$ & $\dsZ$ \\
U(1) \ \ & 0 & $\dsZ$ & 0 & $\dsZ$ & 0 & $\dsZ$ & 0 & $\dsZ$ & 0 & $\dsZ$ \\
SU(2) \ \ & 0 & $\dsZ$ & 0 & $\dsZ_2$ & $\dsZ_2$ & $\dsZ$ & 0 & 0 & 0 & $\dsZ$ \\
\hline\hline
\multicolumn{11}{c}{Boundary anomaly}\\
$d-1$ &  & 1 & 2 & 3 & 4 & 5 & 6 & 7 & 8 & 9\\
\hline
Grav. &  & P &   &   &   & P &   & G & G & P \\
U(1) &  & P &   & P &   & P &   & P &   & P \\
SU(2) &  & P &   & G & G & P &   &   &   & P\\
\hline
\end{tabular}
\end{center}
\label{tab: anomaly}
\end{table}%

\section{2. Example with perturbative anomaly}

In this section we will discuss the most classic system with
perturbative gauge anomaly: the $(3+1)d$ chiral fermion with a
chiral U(1) global symmetry. The Hamiltonian of this system reads
\beqn H = \int d^3x \ \psi^\dagger (\ii \vect{\sigma} \cdot
\vect{\partial}) \psi, \label{3dchiral}\eeqn with an exact U(1)
symmetry $\psi \rightarrow e^{\ii\theta} \psi$, this Hamiltonian can
never be regularized as a $3d$ system, it must be a boundary of a
$4d$ integer quantum Hall state. Notice that this exact U(1)
symmetry was an important assumption in the famous no-go theorem
proved in Ref.~\onlinecite{doublingA,doublingB}. If we couple
\eqnref{3dchiral} to a dynamical U(1) gauge field $A_\mu$, then
the gauge current would be anomalous: \beqn
\partial_\mu j_\mu \sim \epsilon_{\mu\nu\rho\tau} F_{\mu\nu}
F_{\rho \tau}. \eeqn This anomaly means that we cannot view the
boundary as an independent quantum system. Once we take the entire
system into account, gauge anomalies from two opposite boundaries
will cancel each other.

This gauge anomaly is ``perturbative", in the sense that it can be
computed by standard perturbation theory. Based on the argument
from the introduction, \eqnref{3dchiral} cannot be gapped out
without breaking the U(1) symmetry. We can try gapping out
\eqnref{3dchiral} following the same strategy as
Ref.~\onlinecite{TI_fidkowski2,TI_qi,TI_senthil,TI_max,senthilhe3},
namely we first break the U(1) symmetry and gap out
\eqnref{3dchiral}, then try to restore the U(1) symmetry by
proliferating/condensing the topological defects of the U(1) order
parameter.
Ref.~\onlinecite{TI_fidkowski2,TI_qi,TI_senthil,TI_max,senthilhe3}
discussed how to drive the boundary of $3d$ topological insulator
to a topological order, starting with the superconductor phase of
the boundary which spontaneously breaks the U(1) symmetry. At the
boundary of $3d$ topological insulator, the simplest vortex that
can condense has winding number 4 (a strength 4 vortex), and after
condensation the boundary is driven into a fully gapped
topological state.

In our current case, the complex U(1) order parameter $\phi$ would
couple to the fermions in the following way: \beqn \phi
\;\psi^\intercal \ii \sigma^y \psi + H.c. \eeqn In $3d$ space, the
topological defect of a U(1) order parameter is its vortex loop,
and in principle after condensing the vortex loops the U(1) global
symmetry will be restored. However, the resultant state may or may
not be a fully gapped state, depending on the spectrum of the
vortex loop. If the vortex loop itself is gapless, then the
condensate of the vortex loop cannot be a gapped state. In our
current case, the vortex loop has a $1d$ chiral fermion, which
cannot be gapped out at all by any interaction. Thus the way to
gap out the boundary states in
Ref.~\onlinecite{TI_fidkowski2,TI_qi,TI_senthil,TI_max,senthilhe3}
fails in this situation.

The same conclusion holds for arbitrary copies of
\eqnref{3dchiral}, and for vortex loops with arbitrary strength
(winding number). For example, for a single chiral fermion, a
strength$-N$ vortex loop has $1d$ chiral fermions with chiral
central charge $c = N$, which still cannot be gapped out at all.
Thus our conclusion is that \eqnref{3dchiral} with U(1) anomaly
can never be gapped out even by topological order.

\section{3. $3d$ Topological Order with Witten's anomaly}

\subsection{Physical consequence of Witten's anomaly}

The most well-known example of global anomaly, is the SU(2) global
anomaly of $(3+1)d$ chiral fermions discovered by Witten
\cite{wittensu2}. Let us consider a $(3+1)d$ chiral fermion which
forms a fundamental representation of SU(2): \beqn H = \int d^3x \
\sum_{a = 1}^2 \psi^\dagger_a \ii \vect{\sigma} \cdot
\vect{\partial} \psi_a + \cdots , \label{su2}\eeqn under SU(2)
transformation, $\psi_a \rightarrow \exp(\ii \vect{\tau} \cdot
\vect{\theta}/2)_{ab} \psi_b$. In order to guarantee the chemical
potential locates right at the Dirac point, we assume an extra
inversion combined with particle-hole symmetry on the system:
\beqn \mathcal{IC} : \psi \rightarrow \sigma^y \tau^y
\psi^\dagger, \ \ \ \vect{r} \rightarrow - \vect{r}. \eeqn This
symmetry commutes with the global SU(2). After we couple this
system to a dynamical SU(2) gauge field, then there is a large
gauge transformation that changes the sign of the partition
function~\cite{wittensu2}, which implies that the total partition
function of \eqnref{su2} vanishes after considering all the gauge
sectors. This anomaly comes from the mathematical fact that
$\pi_4[S^3] = Z_2$, and it implies that {\it even without the
SU(2) gauge field}, odd copies of $(3+1)d$ chiral fermions with
exact SU(2) global symmetry cannot be realized in $3d$ space, it
must be the boundary of a $4d$ system.

A $4d$ topological superconductor with SU(2) symmetry has $\dsZ_2$
classification, namely for a single copy of \eqnref{su2}, without
breaking the SU(2) symmetry, the system cannot be gapped out at
all; while two copies of \eqnref{su2} can be trivially gapped out
without breaking SU(2) symmetry (see appendix A for details). Our
goal is to study whether we can gap out one single copy of
\eqnref{su2} by driving the system into a topological order,
without breaking any symmetry. In order to do this, we should
first make sure the system has no perturbative anomaly. Thus
\eqnref{su2} should {\it not} have an extra U(1) symmetry $\psi_a
\rightarrow e^{\ii\theta} \psi_a$. The apparent U(1) symmetry of
\eqnref{su2} is merely a low energy emergent phenomenon, or in
other words, the U(1) symmetry must be explicitly broken by the
lattice model in the bulk, thus rigorously speaking the bulk state
must be a topological superconductor rather than a topological
insulator.

What is the physical meaning of the SU(2) Witten anomaly? If we
view gauge transformation $U(x, \tau)$ as an evolution from $\tau
= - \infty$ to $+\infty$, then the space-time configuration of the
trouble-making large gauge transformation $U(x, \tau)$ corresponds
to first creating a pair of SU(2) soliton and anti-soliton pair in
space (the existence of SU(2) soliton is due to the fact that
$\pi_3[SU(2)] = Z$), then rotating the soliton by $2\pi$, and
eventually annihilating the pair. Now let us couple the fermion
$\psi$ to a SU(2) vector $\vect{n}$: \beqn \vect{n} \cdot
\mathrm{Re}[\psi^\intercal \sigma^y \otimes \tau^y \vect{\tau} \psi
],\eeqn The large gauge transformation $U(x, \tau)$ can be
translated into a space-time configuration of $\vect{n}(x, \tau)$:
the process that causes the partition function to change sign,
corresponds to first creating a pair of Hopf soliton and
anti-soliton pair of $\vect{n}$ in space (the existence of Hopf
soliton of $\vect{n}$ is due to the fact $\pi_3[S^2] = Z$), then
rotating the Hopf soliton by $2\pi$, and eventually annihilating
the pair (more detail about this process is explained in the
appendix B). This interpretation of SU(2) anomaly using Hopf soliton
of $\vect{n}$ is equivalent to Witten's interpretation.

\begin{figure}[t]
\begin{center}
\includegraphics[width=200pt]{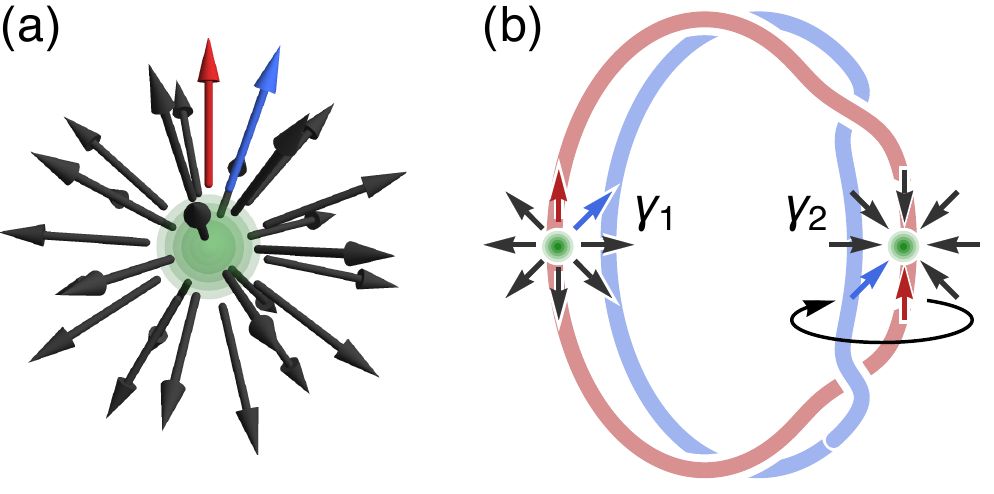}
\caption{(a) A hedgehog monopole traps a Majorana zero mode at its
core. To keep track of the framing, the $\vect{n}=(0,0,1)$ vector
and a nearby vector which is deviated along a tangent direction
are colored in red and blue respectively. (b) The Hopf soliton is
created by an event which corresponds to twisting a hedgehog
monopole in the $3d$ space, i.e. the red and the blue vectors
trace out the two edges of a self-twisted ribbon.} \label{fig:
monopole}
\end{center}
\end{figure}

The fact that Hopf soliton changes sign under $2\pi$ rotation,
implies that Hopf soliton is a fermion. How do we understand the
fermion carried by the Hopf soliton? This was answered in
Ref.~\onlinecite{teokane,ranhopf}. To create a Hopf soliton from
vacuum, we can first create a pair of hedgehog monopole
anti-monopole pair of $\vect{n}$, then rotate the monopole by
$2\pi$, and annihilate the pair, as illustrated in \figref{fig:
monopole}. The final configuration of $\vect{n}$ at $\tau = +
\infty$ compared with the initial configuration at $\tau = -
\infty$ has one extra Hopf soliton. It is well-known that a
hedgehog monopole of $\vect{n}$ has a Majorana fermion zero mode
$\gamma$ localized at the core of the monopole. A pair of
well-separated monopole anti-monopole defines two different
quantum states with opposite fermion parity: $ (-1)^{N_f} =
2\ii\gamma_1 \gamma_2 = \pm 1$. If the monopole anti-monopole pair
has a finite distance, then after rotating the monopole by $2\pi$,
there will be a level crossing in the fermion spectrum, which
causes change of fermion parity of the ground state of the
system~\cite{ranhopf}. This analysis explains why the Hopf soliton
carries a fermion, and also explains the physical meaning of the
Witten's anomaly.

\begin{figure}[htbp]
\begin{center}
\includegraphics[width=140pt]{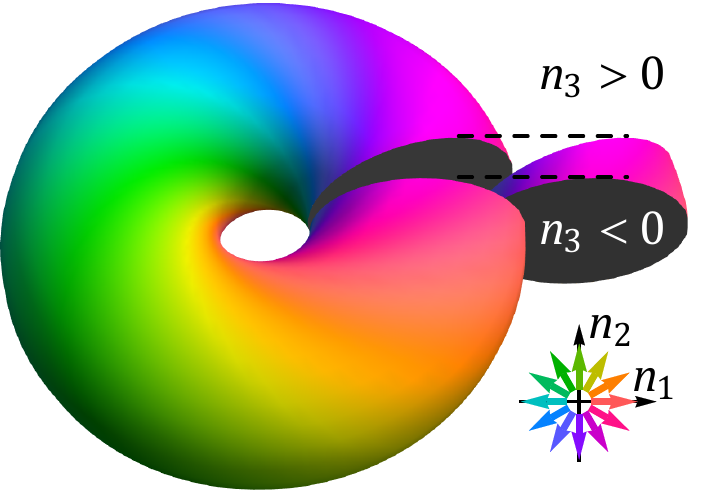}
\caption{Hopf soliton in the $Z_2$ anisotropic limit can be
considered as a torus shape domain wall of $n_3$, on which
$(n_1+\ii n_2)$ (color coded) winds by $2\pi$ around both the
meridian and the longitudinal circles of the torus. A slice of the
torus is cut out to show the sign change of $n_3$ across the
interior and the exterior of the torus. This configuration can
also be viewed as a link of two vortex loops with opposite sign of
$n_3$ respectively.} \label{fig: Hopf Z2}
\end{center}
\end{figure}

The above physical interpretation of Witten's anomaly implies that
this global anomaly also exists in systems whose symmetry is a
subgroup of SU(2), as long as the system still has hedgehog
monopole defect, and the defect carries a Majorana fermion zero
mode. For the convenience of later analysis, let us consider
\eqnref{su2} with $U(1) \rtimes Z_2$ symmetry (a rotation around
$\hat{z}$ axis and $\pi-$rotation around $\hat{x}$ axis), which is
a subgroup of SO(3): \beqn U(1) &:& \psi \rightarrow e^{\ii \tau^z
\theta/2} \psi, \ \ (n_1 + \ii n_2) \rightarrow e^{\ii\theta}(n_1
+ \ii n_2) \cr\cr \mathrm{R}_{x,\pi} &:& \psi \rightarrow
\ii\tau^x \psi, \ \ (n_1, n_2, n_3) \rightarrow (n_1, -n_2, -n_3).
\eeqn With this $U(1) \rtimes Z_2$ symmetry, the classification of
the $4d$ bulk topological superconductor is unchanged, namely
\eqnref{su2} with this reduced U(1)$\rtimes Z_2$ symmetry still
cannot exist in $3d$ space. But with this $U(1) \rtimes Z_2$
symmetry, a hedgehog monopole of $\vect{n}$ becomes a domain wall
of Ising order parameter $n_3$ inside a vortex line of U(1) order
parameter $n_1 + \ii n_2$, and it indeed still carries a Majorana
zero mode. The Hopf soliton, though also distorted compared with
the SU(2) invariant case (see \figref{fig: Hopf Z2}), must still
be a fermion.

In Ref.~\onlinecite{teokane,ranhopf}, because there is no such
$Z_2$ symmetry which transforms $n_3 > 0$ to $n_3 < 0$ ($n_3$ in
our case corresponds to the mass term of bulk Dirac fermion in
Ref.~\onlinecite{teokane,ranhopf}, and the system always polarizes
$n_3$ to be either $n_3
> 0$ or $n_3 < 0$, except for the bulk quantum critical point
between TI and trivial insulator), the system discussed therein
can exist in $3d$, and it is precisely the ordinary $3d$
topological insulator. In $3d$ TI, the hedgehog monopole and Hopf
soliton are always confined because $n_3$ is always polarized;
while in our case, these defects can be deconfined, and this is a
key difference between our $3d$ boundary system and the $3d$ bulk
system in Ref.~\onlinecite{teokane,ranhopf}.

\subsection{$\mathbb{Z}_{2N}$ topological order}

Now let us first gap out \eqnref{su2} by condensing a superfluid
order parameter $n_1 + in_2$, then try to restore the U(1)
symmetry by condensing the vortex loops. Inside the vortex loop,
if $n_3 = 0$, there will be a $1d$ counter propagating nonchiral
gapless Majorana fermion localized in the vortex loop. A nonzero
$n_3$ will open up a gap for this localized modes, and lower the
energy of the vortex loop. Thus energetically the system favors
$n_3$ to be nonzero in the vortex loop. Because $n_3$ can take
either positive or negative expectation values, thus there are two
flavors of vortex loops, whose domain wall is the hedgehog
monopole of $\vect{n}$. With nonzero $n_3$ inside the vortex loop,
the fermion spectrum remains fully gapped. The only potential low
energy fermion excitations are localized inside the hedgehog
monopole (domain of $n_3$ in a vortex loop), but in this work we
will always keep the hedgehog monopole either gapped or confined.
The Hopf soliton now becomes a {\it link} between the two flavors
of vortex loops (see \figref{fig: Hopf Z2}). Because of nonzero
$n_3$ in the vortex loop, this vortex link is a nonsingular smooth
configuration of vector $\vect{n}$.

Let us tentatively ignore the background fermions and the Witten's
anomaly. Using the standard dual description of superfluid in
$(3+1)d$, we can describe these two vortex loops by two gauge
fields $b_{1,\mu}$ and $b_{2,\mu}$. The effective $4d$ Euclidean
space-time theory for vortex loops read~\cite{senthilloop}:
\beqn\label{eq: loop} \mathcal{S} &=& \sum_{\vect{x}}
\sum_{\mu\nu} \sum_{c = 1,2} - t \cos (\nabla_\mu b_{c, \nu} -
\nabla_\nu b_{c, \mu} - 2\pi B_{\mu\nu}) \cr\cr &+& \frac{1}{K}
(\epsilon_{\nu\rho\tau} \nabla_\nu B_{\rho\tau} )^2. \label{dual}
\eeqn The sum is taken over all space-time position $\vect{x}$ and
plaquettes. $B_{\mu\nu}$ is a rank-2 antisymmetric tensor field,
which is the dual of the Goldstone mode of the order parameter
$(n_1 +\ii n_2)$. $\Psi_{c, \mu}^\dagger \sim \exp(\ii b_{c,\mu})$
creates a segment of vortex loop with flavor $c$ along the direction
$\mu$, and $\exp (\ii \nabla \times b)$ creates a small vortex loop.

The pure bosonic theory \eqnref{dual} can have the following
different phases:

{\it (1)} Ordinary superfluid phase. This phase corresponds to the
case when all loops are small and gapped. Then the system has only
one gapless mode described by $B_{\mu\nu}$.

{\it (2)} U(1) liquid phase with gapless photon excitation, which
was discussed in Ref.~\onlinecite{senthilloop}. This phase
corresponds to the case when all monopoles are gapped, while both
flavors of vortex loops $b_{1, \mu}$ and $b_{2, \mu}$ condense. In
this phase, the linear combination $b_{1, \mu} + b_{2, \mu}$
(which corresponds to the bound state between the two flavors of
vortices $\Psi^\dagger_{1,\mu} \Psi^\dagger_{2,\mu}$) will
``Higgs" and gap out $B_{\mu\nu}$, and the combination $b_{1, \mu}
- b_{2, \mu}$ becomes the gapless photon mode of the U(1) liquid.

{\it (3)} A {\it fully gapped} $\mathbb{Z}_{2N}$ topological order
which preserves all the symmetries. This is a phase where
individual loop $b_{1,\mu}$ and $b_{2,\mu}$ does {\it not}
condense, but vortex bound state $(\Psi^\dagger_{1,\mu}
\Psi^\dagger_{2,\mu})^N \sim \exp (\ii N b_{1,\mu} + \ii N
b_{2,\mu}) = \exp(\ii b_\mu)$ condense and gap out $B_{\mu\nu}$
through Higgs mechanism. Because now $b_\mu$ is a bound state of
$2N$ vortex loops, this phase is a $\mathbb{Z}_{2N}$ topological
order. It is well-known that condensation of double vortex loops
will lead to a $\mathbb{Z}_2$ topological order with
fractionalization, for instance see
Ref.~\onlinecite{doublevortex1,doublevortex2,doublevortex3}. A
condensate of $2N$ vortex loop bound state can be effectively
described by the following action: \beqn \label{Z2N} \mathcal{S}
&=& \sum_{\vect{x}} \sum_{\mu\nu} - t \cos (\nabla_\mu b_\nu -
\nabla_\nu b_{\mu} - 2\pi (2 N B_{\mu\nu})) \cr\cr &+& \frac{1}{K}
(\epsilon_{\nu\rho\tau} \nabla_\nu B_{\rho\tau} )^2. \eeqn It is
clear that when $b_\mu$ condenses, $B_{\mu\nu}$ takes only $2N$
discrete values $0, \frac{1}{2N}, \cdots \frac{2N-1}{2N}$, hence
the condensate is a $\mathbb{Z}_{2N}$ topological order. Also,
under the $Z_2$ symmetry transformation, $\Psi_{1,2} \rightarrow
\Psi_{2,1}^\dagger$, $i.e.$ $b \rightarrow - b$, the vortex loop
condensate explicitly preserves the $Z_2$ symmetry as long as we
take $b = 0$ in the condensate.

The topological order {\it (3)} is what we will focus on in this
paper. In our case, because of the background fermions and the
Witten's anomaly, there is one subtlety that we need to be careful
with. With odd $N$, say $N = 1$, a bound state $\Psi_1 \Psi_2$
could be a fermion if $\Psi_1$ and $\Psi_2$ has odd number of
links in the space, due to the Witten's anomaly. Thus we should
condense only the configurations of vortex loop bound state in
which $\Psi_1$ and $\Psi_2$ are always parallel and properly
separately so that they are not linked at all. We assume this can
be achieved by turning on local interactions between the loops,
although we do not prove this. For even integer $N$ this subtlety
does not arise at all, because the link between $(\Psi_1)^{N}$ and
$(\Psi_2)^{N}$ is always a boson, thus their bound state is free
to condense. In the following we will take $N = 1$ as an example
($\mathbb{Z}_2$ topological order), but our discussion can be
generalized to arbitrary integer $N$.

An ordinary $\mathbb{Z}_2$ topological order can be driven into a
trivial gapped confined phase by proliferating/condensing the
``vison loops". A vison loop in our case is bound with a single
vortex loop of order parameter $n_1 + in_2$. In the following we
will argue that our $\mathbb{Z}_2$ topological order is a special
one, it {\it cannot} be further driven into a trivial confined
phase.

First of all, we still have two flavors of vison loops with fully
gapped fermion spectrum, which corresponds to $n_3 > 0$ or $n_3 <
0$ at the vortex core. We will primarily consider the vison loops
with uniform $n_3$, or in other words the vison loops in which the
$Z_2$ symmetry and the $\mathcal{IC}$ symmetry are spontaneously
broken. This is because if inside the vison loop there is a domain
wall of $n_3$, at the domain wall there will be a Majorana fermion
zero mode, and it is unclear whether these vison loops with
Majorana zero modes can condense at all due to the non-Abelian
statistics introduced by the Majorana zero modes. And if a vison
loop has $\langle n_3 \rangle = 0$, then the fermions will be
gapless along the vison loop, and condensing these vison loops
will not lead to a trivial gapped confined phase.

If we do not want to break any symmetry, the two flavors of fully
gapped vison loops must condense simultaneously. However, there is
a clear obstacle for condensing both vison loops like an ordinary
$\mathbb{Z}_2$ topological order. This is because when these two
different vison loops are linked, the $\vect{n}$ configuration
around the vison link is a Hopf soliton, and hence it must be a
fermion, which is a consequence of the Witten's anomaly. By
contrast, if the $Z_2$ symmetry is explicitly broken (for instance
in the $3d$ TI), then we can condense just one flavor of vison
loops while preserving all the symmetries, then it is possible to
get a fully symmetric confined phase.

The hedgehog monopole of the superfluid phase becomes the end
point of the loop $b_{1,\mu} - b_{2,\mu}$. While because this loop
does not condense in the $\mathbb{Z}_2$ topological order, this
loop still has a finite loop tension, hence the hedgehog monopole
which carries Majorana fermion zero mode is still {\it confined}
in the $\mathbb{Z}_2$ topological order. In
Ref.~\onlinecite{qinayak} the authors discussed a gapless phase
where the hedgehog monopole is deconfined. Whether there is a
fully gapped topological phase with deconfined hedgehog monopole
which carries Majorana fermion zero mode is an open question.

\subsection{CP$^1$ formalism}

All we have discussed so far can be equivalently formulated in the
standard CP$^1$ formalism, which was also used in
Ref.~\onlinecite{qinayak} to study the gapless photon phase. In
the CP$^1$ formalism, the order parameter $\vect{n}$ is
fractionalized into the bosonic spinon $z=(z_1,z_2)^\intercal$ via
$\vect{n}=z^\dagger\vect{\tau}z$ under the constraint of
$z^\dagger z=1$. The constraint can be implemented by the emergent
U(1) gauge field $a_\mu$ between the spinons. The symmetry acts on
the spinon as $\text{U(1)}: z\to \exp(\ii\tau^z\theta/2)z$ and
$\mathrm{R}_{x,\pi}: z \to \tau^x z$. The field theory for both
the CP$^1$ spinon and the SU(2) chiral fermion on the $3d$
boundary reads\cite{qinayak} \beq\label{eq: CP1}
\begin{split}
\mathcal{S}=&\frac{1}{2g}|(\ii\partial-a )_\mu z|^2+\mu(z^\dagger z-1)\\
&+\psi^\dagger(\ii\partial_0+\ii\vect{\sigma}\cdot\vect{\partial})
\psi+z^\dagger\vect{\tau}z\cdot\Re[\psi^\intercal\sigma^y\tau^y\vect{\tau}\psi].
\end{split}
\eeq The spinons $z_1$ and $z_2$ carry $\pm1/2$ U(1) symmetry
charges respectively, and both carry one U(1) gauge charge.

Let us start with the ordered phase of $\vect{n} =
z^\dagger\vect{\tau}z$, \emph{i.e.} the spinon condensed phase
$\langle z\rangle\neq 0$. In this phase, the chiral fermion $\psi$
is fully gapped, the gauge U(1) fluctuation $a_\mu$ is Higgsed out
by the $z$ condensate, and the symmetry U(1) is spontaneously
broken leading to one gapless Goldstone mode of $n_1 + \ii n_2
\simeq z_1^*z_2$.

We consider the Hopf soliton configuration of $\vect{n}$, which is
also a pair of $n_1 + \ii n_2$ vortices ($2\pi$ symmetry fluxes)
linked together. Each vortex must be bound to a $\pi$ gauge flux
of $a_\mu$ to reduce the kinetic energy of the spinon, thus the
Hopf soliton corresponds to a linking of $\pi$ gauge fluxes whose
linking number is counted by the Chern-Simon term as
$\frac{1}{\pi^2}\int a \wedge \dd a = 1$ (see Appendix B the
correspondence of Hopf soliton and gauge flux link). Suppose the
typical length scale of the Hopf soliton is $R$, to preserve the
linking number given by the Chern-Simon term, $a$ must scale with
$R$ as $a \sim 1/R$, so the Maxwell term of the U(1) gauge field
will contribute energy $E \simeq \int \frac{\kappa}{2}(\dd a)^2
\sim \kappa /R$. In the ordered phase, the condensate of $z_1$ and
$z_2$ will generate a mass term $a^2$ to the effective action,
then the soliton energy is given by $E \simeq \int
\frac{\kappa}{2} (\dd a)^2 + \frac{\rho}{2} a^2$, which scales
with $R$ as $E \sim \kappa/R+\rho R$ and is minimized at at finite
length scale $R_0\sim(\kappa/\rho)^{1/2}$ with a finite energy
$E_0\sim(\kappa\rho)^{1/2}$. In the $\mathbb{Z}_2$ topological
order we discussed in the last section, although $z_1$ and $z_2$
are not individually condensed, the boundary state $z_1z_2$ is
still condensed which breaks the U(1) gauge field down to
$\dsZ_2$, and a mass term $a^2$ still exists for the gauge field.
Thus the Hopf soliton becomes a local object and can be fully
gapped out. Since the spinon pair $z_1z_2$ carries two units of
gauge charge and no symmetry charge, the U(1) gauge structure is
broken down to $\dsZ_2$ without breaking any physical symmetry,
therefore we obtain a fully gapped symmetric $\mathbb{Z}_2$
topological order on the $3d$ boundary.

To make connection to the loop theory in \eqnref{eq: loop}, we
evoke the duality transformation. To start, we rewrite the CP$^1$
field $z_c  \sim e^{\ii\theta_c}$ ($c = 1,2$) in terms of the
phase angles $\theta_c$. We can neglect the amplitude fluctuation
of each $z_c$ component, as long as we take the easy-plane limit
of the system, $i.e.$ $n_1$ and $n_2$ are energetically more
favorable than $n_3$. In this limit, the effective action in the
Euclidean space-time reads \beq \mathcal{S} = \sum_{c=1,2} -
K\cos(\dd \theta_c-a) \eeq We can take the standard Villain form
of the action, by expanding the cosine function at its minimum,
and introducing the 1-form fields $l_c\in\dsZ$ and
$k_c\in\mathbb{R}$ ($c=1,2$): \beq
\begin{split}
\mathcal{Z}=&\Tr \exp\Big[\sum_{c=1,2} -\frac{K}{2}(\dd \theta_c-a-2\pi l_c)^2\Big]\\
\sim&\Tr \exp\Big[\sum_{c=1,2} \frac{1}{2K}k_c^2+k_c\cdot(\dd \theta_c-a-2\pi l_c)\Big]\\
\sim&\Tr \exp\Big[\sum_{c=1,2} \frac{1}{2K}k_c^2-k_c\cdot(a+2\pi l_c)\Big]\delta[\partial k_c]\\
\sim&\Tr \exp\Big[\sum_{c=1,2} \frac{1}{2K}(\dd B_c)^2+(a+2\pi l_c)\wedge \dd B_c\Big].
\end{split}
\eeq In the last line, we introduce the 2-form fields $B_c$
($c=1,2$) on the dual space-time manifold, such that $k_c=\star\dd
B_c$ resolves the constraint $\partial k_c=0$. Summing over $l_c$
will require $B_c$ to take only integer values, which could be
imposed by adding a $\cos(2\pi B_c)$ term, and the theory now
becomes \beq \mathcal{Z}\sim\Tr \exp\Big[\sum_{c=1,2}
\frac{1}{2K}(\dd B_c)^2+a\wedge \dd B_c-t\cos(2\pi B_c)\Big] \eeq
Integrating out the gauge field $a$ will impose the constraint
$\dd(B_1 + B_2)=0$, which can be resolved by $B_1 = B - \dd
b_1/(2\pi)$, $B_2 = -B + \dd b_2/(2\pi)$. Therefore the final
action takes the form of $\mathcal{S} \sim \sum_{c = 1,2} - t
\cos(\dd b_c - 2\pi B) + K^{-1}(\dd B)^2$ which is identical to
\eqnref{dual}.

$b_{1,\mu}$ and $b_{2,\mu}$ introduced in \eqnref{dual} correspond
to vortex of $z_1$ and anti-vortex of $z_2$ respectively, which
both correspond to a vortex of the original order parameter $n_1 +
\ii n_2 \sim z_1^\ast z_2$. condensation of the vortex bound state
$b_{1,\mu} + b_{2,\mu}$ in \eqnref{dual} will disorder the
physical order parameter $z_1^\ast z_2$, but not disorder the
condensate $z_1 z_2$, thus the $\mathbb{Z}_2$ topological state
after condensing $b_{1,\mu} + b_{2,\mu}$ is precisely the same
$\mathbb{Z}_2$ topological state after condensation of bound state
$z_1 z_2$ in the CP$^1$ formalism. And the loop excitations
$b_{1,\mu}$ and $b_{2,\mu}$ of this $Z_2$ topological state both
correspond to the $\pi-$flux lines of $a_\mu$.

Since our $\mathbb{Z}_2$ topological order is obtained by
condensing pair of $z_1 z_2$ from the U(1) photon phase, the
hedgehog monopole of $\vect{n}$, which in the U(1) photon phase
becomes the Dirac monopole of $a_\mu$, will be {\it confined} in
our $\mathbb{Z}_2$ topological order. The remnant of the Witten's
anomaly is completely encoded in the fact that the link of two
flavors of vison loops must be a fermion, thus the $\mathbb{Z}_2$
topological order cannot be driven into a trivial confined phase
that preserves all the symmetries.

As we discussed in the last section, if we start with the
superfluid phase and condense vortex bound state $(\Psi_1
\Psi_2)^N$, the system will enter a $\mathbb{Z}_{2N}$ topological
order. Starting with the CP$^1$ formalism, this $\mathbb{Z}_{2N}$
topological order can be understood as following: the condensate
$z_1 z_2$ implies that $z_1 \sim z_2^\ast$. The $Z_{2N}$ gauge
field is introduced by fractionalizing the CP$^1$ field as $z_1
\sim z_2^\ast \sim w^N$ with bosonic parton field $w$.
Equivalently we can write $n_1 - \ii n_2 \sim z_1 z_2^\ast \sim
w^{2N}$. Because $z_1 \sim z_2^\ast$ carries U(1) charge $1/2$,
the parton $w$ carries global U(1) charge $1/(2N)$, which is
consistent with the $2N$ flux condensate. The parton $w$ is
coupled to the $Z_{2N}$ gauge field.

\section{4. $4d$ Topological Order with Global Anomaly}

Analogue of Witten's anomaly can be found in higher dimensions.
The simplest generalization is in one higher dimension: one single
copy of $(4+1)d$ Dirac fermion with SU(2) or U(1)$\rtimes Z_2$
symmetry cannot exist in $(4+1)d$ itself, it must be a boundary of
a $5d$ topological superconductor: \beqn H = \int d^4x \
\psi^\dagger (\ii \vect{\Gamma} \cdot \vect{\partial}) \psi \eeqn
where $\vect{\Gamma}=(\Gamma^1,\Gamma^2,\Gamma^3,\Gamma^4)$ and we
choose $\Gamma^{1,2,3}=\sigma^{3}\otimes\sigma^{1,2,3}$,
$\Gamma^{4,5}=\sigma^{1,2}\otimes\sigma^0$. The fermions transform
under the symmetry as U(1): $\psi\to e^{\ii \theta/2}\psi$ and
$\mathrm{R}_{x,\pi}: \psi\to \ii\Gamma^5\Gamma^2\psi^\dagger$. Now
we couple the fermion $\psi$ to a vector $\vect{n}$ as \beq
(n_1-\ii n_2)\psi^\intercal \Gamma^2 \psi  +n_3 \psi^\dagger
\Gamma^5 \psi + H.c. \eeq such that the vector $\vect{n}$
transforms as U(1): $(n_1+\ii n_2)\to e^{\ii\theta}(n_1+\ii n_2)$
and $\mathrm{R}_{x,\pi}: n_{2,3}\to-n_{2,3}$.

\begin{figure}[htbp]
\begin{center}
\includegraphics[width=160pt]{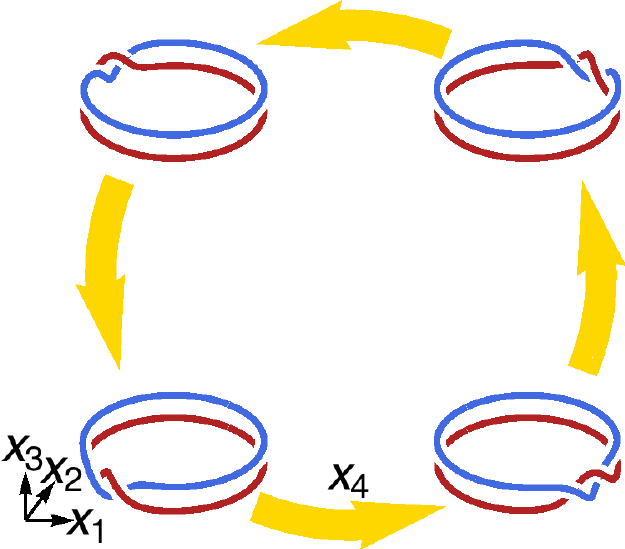}
\caption{$\pi_4[S^2]$ soliton in the $4d$ space. As one circles
around the 4th dimension $x_4$, the Hopf soliton in each $3d$
slice rotates around. The Hopf soliton is represented as linked
preimages of $n_3=\pm1$ points on $S^2$. The configuration also
corresponds to two $2d$ gauge flux membranes linked in the $4d$
space, s.t. in each $3d$ slice, the sections of flux membranes are
linked flux loops (red and blue loops).} \label{fig: soliton}
\end{center}
\end{figure}

In $4d$ space vector $\vect{n}$ also has a nontrivial soliton. The
$\pi_4[S^2]$ soliton configuration is given by the non-trivial map
$f:S^4\to S^2$, which can be considered as the composition of two
non-trivial maps $g:S^4\to S^3$ and $h:S^3\to S^2$ as $f=h\circ
g$. The first map $g$ is such that the preimage of each point in
$S^3$ is a circle in $S^4$, along which the $3d$ framing twists
around once. The second map $h$ is just the standard Hopf map. So
the $\pi_4[S^2]$ soliton can be understood by considering $3d$
slices embedded in the $4d$ space, with each slice hosting a Hopf
soliton, and the Hopf soliton rotates by $2\pi$ as the slice
evolves along the 4th dimension. Also, while mapping $4d$ space to
$S^2$, every preimage of $S^2$ is a $2d$ manifold (for instance
$T^2$ or $S^2$). And two disconnected $m-$dimensional manifolds
can have nontrivial linking in $(m+2)-$dimensional space (knot
with codimension-2). A nontrivial $\pi_4[S^2]$ soliton corresponds
to the case when the preimages of two arbitrary points on $S^2$
will be two $2d$ manifolds linked in the $4d$ space.

\begin{figure}[t]
\begin{center}
\includegraphics[width=150pt]{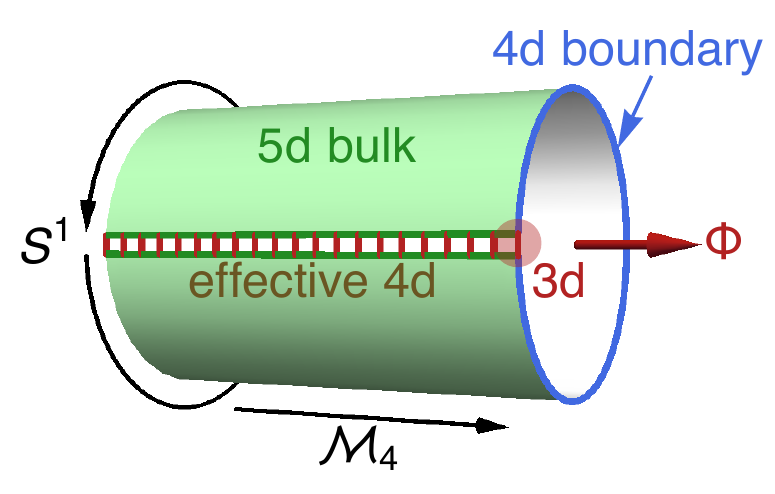}
\caption{Compactfy the $5d$ bulk with a flux $\Phi$ through the
compactified dimension $S^1$.} \label{fig: 5d}
\end{center}
\end{figure}

Now we argue that the $\pi_4[S^2]$ soliton on the $4d$ boundary of
the $5d$ topological superconductor is also fermionic. We first
consider the $5d$ bulk as a $\mathcal{M}_4\times S^1$ manifold
(see \figref{fig: 5d}) where $\mathcal{M}_4$ is a $4d$ manifold,
and then compactify the $S^1$ dimension. Depending on the flux
$\Phi$ threaded through the $S^1$,  the compactified effective
$4d$ system can either be a trivial superconductor ($\Phi=0$) or a
topological superconductor ($\Phi=\pi$). This can be shown
explicitly by the cut-and-glue strategy: first cut the $5d$ bulk
along the $\mathcal{M}_4$ to expose the upper and the lower $4d$
boundaries (green boundaries in \figref{fig: 5d}), described by
$H_\text{cut}=\int\dd^4 x\;\psi_1^\dagger
(\ii\vect{\Gamma}\cdot\vect{\partial})\psi_1-\psi_2^\dagger
(\ii\vect{\Gamma}\cdot\vect{\partial})\psi_2$, and then glue the
boundaries together by a coupling term $H_\text{glue}=u
\int\dd^4x\;\ii\psi_1^\dagger\psi_2+H.c.$ with the coupling
coefficient $u\sim e^{\ii \Phi}$ depending on the flux $\Phi$
through the $S^1$. $H_\text{cut}+H_\text{glue}$ together describes
an effective $4d$ superconductor with the U(1)$\rtimes Z_2$
symmetry that U(1): $\psi_a\to e^{\ii\theta/2}\psi_a$ and
$\mathrm{R}_{x,\pi}:\psi_a\to\ii\Gamma^5\Gamma^2\psi_a$ ($a=1,2$).
As the flux $\Phi=0,\pi$: $u=\pm1$ plays the role of the
topological mass that tunes the $4d$ effective bulk state between
the trivial and the topological phases. In the presence of the
$\pi$ flux ($\Phi=\pi$), $H_\text{cut}+H_\text{glue}$ together
describes an effective $4d$ topological superconductor protected
by U(1)$\rtimes Z_2$ whose $3d$ boundary has the Witten anomaly,
namely the Hopf soliton is fermionic on the compactified $3d$
boundary. If we revert the compactification, the original $4d$
boundary (blue boundary in \figref{fig: 5d}) of the $5d$ bulk is
the $S^1$ extension of the $3d$ boundary of the effective $4d$
bulk. The $\pi$ flux of the fermions corresponds to the $2\pi$
vortex of the order parameter $n_1+\ii n_2$. So the configuration
of $\vect{n}$ on the $4d$ boundary is indeed a $3d$ Hopf soliton
rotated by $2\pi$ as it translated around in the $S^1$ dimension,
which corresponds to a $\pi_4[S^2]$ soliton. Thus the fermionic
nature of the Hopf soliton on the $3d$ boundary of the $4d$
topological superconductor implies that the $\pi_4[S^2]$ soliton
on the $4d$ boundary of the $5d$ topological superconductor is
also fermionic.

We can now drive the $4d$ boundary to a $\mathbb{Z}_2$ (or
$\mathbb{Z}_{2N}$) topological order, following the same strategy
of the previous section. We can first condense $(n_1, n_2)$ and
spontaneously break the U(1) symmetry. In $4d$ space, the
topological defects of a superfluid phase are $2d$ vortex
membranes, and there are still two flavors of vortex membranes
$\Psi_{1,\mu\nu}\sim \exp(\ii b_{1, \mu\nu})$ and $\Psi_{2,\mu\nu}
\sim (\ii b_{2,\mu\nu})$ depending on the sign of $n_3$ in the
vortex core. Without the $n_3$ component in the vortex core, the
vortex membrane will host a single $2d$ gapless Majorana cone. In
this case, the vortex condensation will lead to gapless boundary
which is not what we are after. However, once we introduce the
$n_3$ component in the vortex core, the Majorana cone is gapped
out in both $b_1$ membrane and $b_2$ membranes. Then by condensing
vortex membrane bound state $(\Psi_1\Psi_2)^N$, this system is
driven into a $\mathbb{Z}_{2N}$ topological order. Again, in this
topological order, $\Psi_1$ and $\Psi_2$ become two flavors of $2d
$ unit gauge flux membranes, and when they ``link" in $4d$ space,
the configuration of $\vect{n}$ around this link will be a
$\pi_4[S^2]$ soliton, and hence it must be a fermion. Thus this
$\mathbb{Z}_{2N}$ topological order can not be further driven into
a trivial confined phase, unless we explicitly break the $Z_2$
symmetry.

\section{5. Implication and Summary}

\begin{figure}[t]
\begin{center}
\includegraphics[width=180pt]{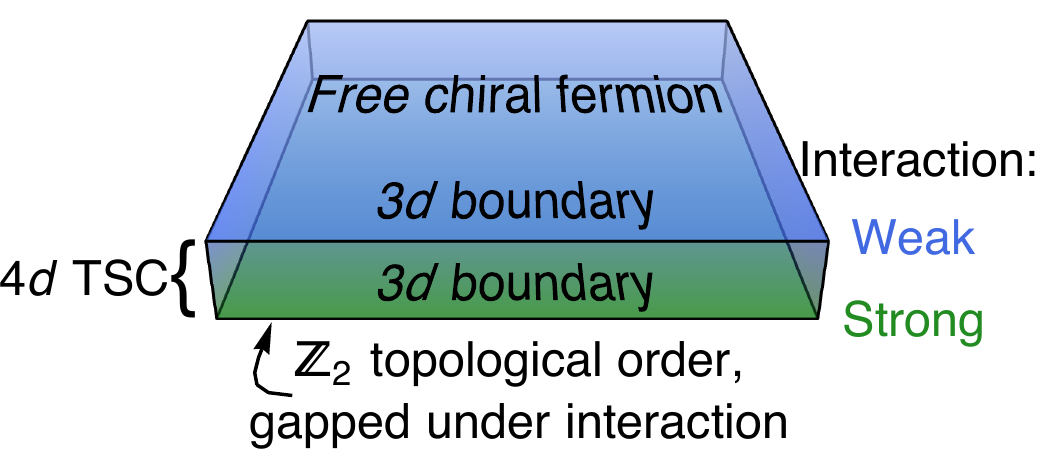}
\caption{A slab of $4d$ topological superconductor with
U(1)$\rtimes Z_2$ symmetry. The interaction strength changes from
weak on the top boundary (in blue) to strong on the bottom
boundary (in green), such that the bottom boundary is gapped out
by the $\dsZ_2$ topological order.} \label{fig: slab}
\end{center}
\end{figure}

Our analysis in this work implies that we can realize some exotic
states in $3d$ systems. For example, let us consider a slab of
$4d$ system, with a thin fourth dimension, as shown in
\figref{fig: slab}. Because the fourth dimension is finite, the
entire system is three dimensional, but we can still realize two
different $3d$ boundary states on two opposite boundaries: the top
boundary is a free chiral fermion \eqnref{su2} with exact
U(1)$\rtimes Z_2$ symmetry, the bottom boundary is the fully
gapped $\dsZ_2$ topological order in which the link of two vison
loops is a fermion. This state is possible as long as we make the
interaction stronger on the bottom boundary, but weaker on the top
boundary. Because a short range interaction on \eqnref{su2} is
irrelevant, \eqnref{su2} will survive at low energy for weak
interaction. But because the bottom boundary is fully gapped, any
low energy experiment can only probe the top surface, which may
lead to the conclusion that this system is ``anomalous". But the
entire system, including both the top and bottom boundary, is
anomaly free.

In the following we will list a few open questions that we were
not able to address in this paper:

{\it i.} In this paper we have understood the topological order
for systems with U(1)$\rtimes Z_2$ symmetry, which still has
Witten's anomaly. However, our formalism in terms of vortex loop
condensation does not directly apply to systems with SU(2)
symmetry, because a precise duality formalism has not been
developed for systems with SU(2) symmetry, thus we have not proved
that our topological order can survive in the SU(2) limit,
although we do not see a fundamental obstacle for that.

{\it ii.} As we explained in this paper, in our topological order,
the nonabelian topological defect which carries the Majorana
fermion zero mode, $i.e.$ the hedgehog monopole of $\vect{n}$, is
still confined. Whether there is a fully gapped topological order
with deconfined $3d$ nonabelian defect is still an open question.


{\it iii.} In this paper we studied the boundary of two
topological superconductors whose boundary states have global
gauge anomaly after ``gauging", and we demonstrated that these two
systems can both be driven into a boundary topological orders. But
if a system involves nonunitary symmetries that cannot be
``gauged", the situation seems to be more complicated. As we
mentioned in the introduction, Ref.~\onlinecite{senthilhe3} has
given us an example of $3d$ topological superconductor whose
boundary can never be driven into a gapped topological order. Thus
a more refined classification of ``gappable" and ``ungappable"
anomalous systems is demanded for systems that involve
time-reversal symmetry.

The authors are supported by the the David and Lucile Packard
Foundation and NSF Grant No. DMR-1151208. The authors thank
Xiao-Gang Wen, Chetan Nayak and Xiao-Liang Qi for very helpful
discussions.

\bibliography{anomaly}

\maketitle \onecolumngrid
\appendix

\section{A. Topological superconductors with SU(2) symmetry}

The topological superconductors/insulators with SU(2) symmetry
belongs to the symmetry class C in the classification
table\cite{ludwigclass1, ludwigclass2} of fermion SPT phases. In
$(4+1)d$ and $(5+1)d$, the SU(2) fermion SPT phases are $\dsZ_2$
classified, and the classification remains the same under the
interaction. In the following we will give a brief introduction to
these topological superconductors/insulators with explicit model
Hamiltonian and symmetry actions.

\subsection{A1. $(4+1)d$ bulk with $(3+1)d$ boundary}
The $4d$ topological superconductor with SU(2) symmetry can be
described by the following lattice model \beq\label{eq: 4dTSC}
H=\sum_{\vect{k}}\sum_{a=1,2}\psi_a^\dagger\Big(\sum_{i=1}^4\sin
k_i \Gamma^i+\Big(\sum_{i=1}^4 \cos
k_i-4+m\Big)\Gamma^5\Big)\psi_a, \eeq where the fermion
$\psi=(\psi_1,\psi_2)$ forms  a fundamental representation of the
SU(2) symmetry: $\psi\to
\exp(\ii\vect{\tau}\cdot\vect{\theta}/2)\psi$. The $\Gamma^i$
matrices are defined as
$\Gamma^{1,2,3}=\sigma^{3}\otimes\sigma^{1,2,3}$,
$\Gamma^4=\sigma^1\otimes\sigma^0$,
$\Gamma^5=\sigma^2\otimes\sigma^0$ with $\sigma^{0,1,2,3}$ being
the Pauli matrices. The model has an emergent U(1) symmetry
$\psi\to e^{\ii\theta}\psi$, which is not required. It is possible
to turn on a weak SU(2)-singlet pairing $(\Delta
\psi^\intercal\tau^y\psi+H.c.)$ to break the U(1) symmetry
explicitly in the bulk, while still retaining the SU(2) symmetry,
hence the system is a superconductor in general. When $m>0$
($m<0$), the model is in its topological (trivial) phase.

The $3d$ boundary of the $4d$ topological superconductor will host
the gapless chiral fermion which carries the SU(2) fundamental
representation. The boundary effective Hamiltonian is \beq
H_\partial=\int\dd^3
x\;\sum_{a=1,2}\psi_a^{\dagger}(\ii\vect{\partial}\cdot\vect{\sigma})\psi_a,
\eeq where $\psi=(\psi_1,\psi_2)^\intercal$ forms the SU(2)
doublet (as inherited from the bulk fermion). The boundary fermion
mode can not be gapped out (on the free fermion level) due to the
SU(2) anomaly. The only possible fermion mass term that can be
added to the boundary theory is the pairing term
$\Delta_{ab}\psi_a^\intercal\ii\sigma^2\psi_b$, however such term
necessarily breaks the SU(2) symmetry. Because the fermion
statistics requires $\Delta_{ab}=\Delta_{ba}$ to be symmetric, so
the pairing term must be an SU(2) triplet, and thus breaks the
symmetry. However if we double the system, then we can gap out the
boundary by introducing the pairing term $\psi^\intercal
\ii\sigma^2\tau^2\mu^2\psi$ where $\tau^i$ and $\mu^i$ are the
Pauli matrices that act in the spaces of the SU(2) spinor and the
two copies of the fermions respectively. Therefore the SU(2)
topological superconductor in $4d$ is $\dsZ_2$ classified. This
classification will not be further reduced by the fermion
interaction.

\subsection{A2. $(5+1)d$ bulk with $(4+1)d$ boundary}
The $5d$ topological insulator with SU(2) symmetry can be
described by the following lattice model \beq H=\sum_{\vect{k}}
\sum_{i=1}^5 \sin k_i \big(\psi_{1,\vect{k}}^\dagger \Gamma^i
\psi_{1,\vect{k}}-\psi_{2,\vect{k}}^\dagger \Gamma^i
\psi_{2,\vect{k}}\big)+ \Big(\sum_{i=1}^5 \cos k_i-5+m\Big)
(\ii\psi_{1,\vect{k}}^\dagger\psi_{2,\vect{k}}+H.c.), \eeq where
$\psi=(\psi_1,\psi_2)^\intercal$ is an 8-component complex fermion
field ($\psi_1$ and $\psi_2$ are both 4-component), and the five
gamma matrices $\Gamma^i$ ($i=1,\cdots,5$) are defined below
\eqnref{eq: 4dTSC}. The model has an SU(2) symmetry, which
transforms $(\psi_{\vect{k}},
\Gamma^2\Gamma^5\psi_{-\vect{k}}^\dagger)^\intercal$ as an SU(2)
doublet (fundamental representation). The obvious U(1) symmetry
$\psi\to e^{\ii\theta}\psi$ is a subgroup of  the SU(2) symmetry.
When $m>0$ ($m<0$), the model is in its topological (trivial)
phase.

The $4d$ boundary of the $5d$ topological insulator will host the
gapless fermion \beq H_\partial=\int\dd^4
x\;\psi^{\dagger}(\ii\vect{\partial}\cdot\vect{\Gamma})\psi, \eeq
where $(\psi,\Gamma^2\Gamma^5\psi^{\dagger})^\intercal$ forms the
SU(2) doublet (as inherited from the bulk fermion). The boundary
fermion mode can not be gapped out (on the free fermion level) due
to the SU(2) anomaly. The only possible fermion mass terms that
can be added to gap out the boundary are $\psi^\intercal \Gamma^2
\psi$ and $\psi^\dagger \Gamma^5 \psi$. It can be verified that
all these mass terms break the SU(2) symmetry. However if we
double the system, then we can gap out the boundary by introducing
the SU(2) symmetric mass term
$\psi_{A}^\dagger\ii\Gamma^5\psi_{B}+H.c.$ where $A,B$ labels the
two copies of the fermions. Therefore the SU(2) topological
superconductor in $5d$ is $\dsZ_2$ classified. This classification
will not be further reduced by the fermion interaction.

\section{B. Topological defects}

\subsection{B1. SU(2) soliton and Hopf soliton}

\begin{figure}[htbp]
\begin{center}
\includegraphics[width=240pt]{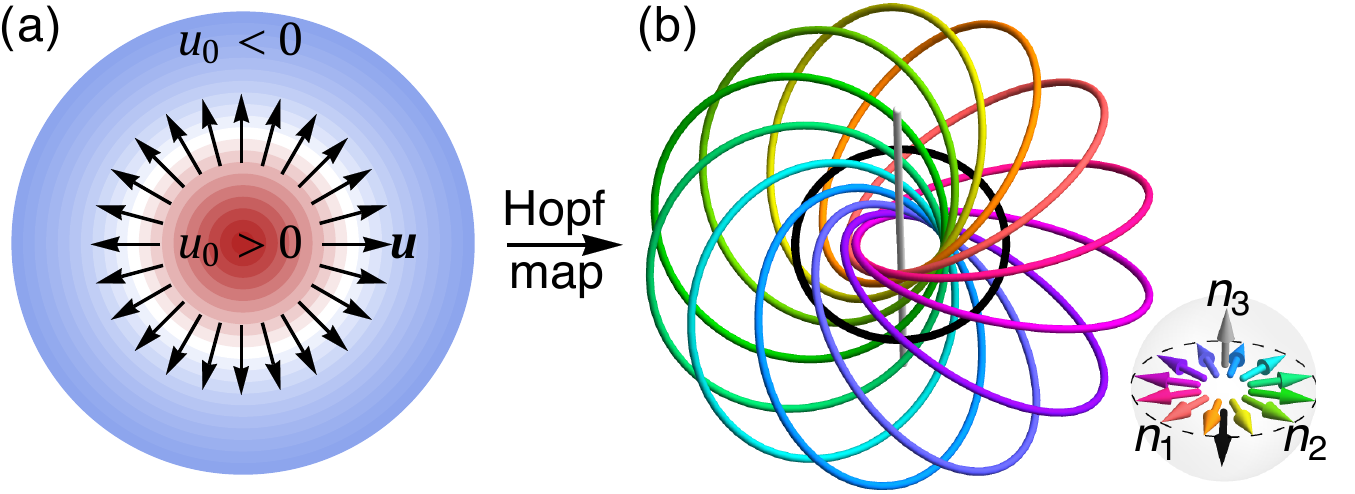}
\caption{(a) SU(2) soliton and (b) Hopf soliton.}
\label{fig: Hopf}
\end{center}
\end{figure}

Consider an SU(2) field $U(\vect{r})$ in the $3d$ space with $U$
being a $2\times2$ unitary matrix, which can be parameterized by
an O(4) vector $\vec{u}=(u_0,u_1,u_2,u_3)\in S^3$ as
\beq\label{eq: U} U=u_0\sigma^0+\ii u_1\sigma^1+\ii
u_2\sigma^2+\ii u_3\sigma^3.\eeq An SU(2) soliton can be given by
the following configuration (in Cartesian coordinate)
\beq\label{eq: u(r)}
\vec{u}(\vect{r})=\frac{1}{1+r^2}(1-r^2,2x,2y,2z), \eeq where
$\vect{r}=(x,y,z)$ and $r=|\vect{r}|$. It is straight forward to
verify that $|\vec{u}(\vect{r})|=1$ through out the space. As
shown in \figref{fig: Hopf}(a), the configuration of $\vec{u}$ is
a hedgehog monopole of $\vect{u}=(u_1,u_2,u_3)\sim\vect{r}$ around
$|\vect{r}|\sim1$, with its interior filled by $u_0\to+1$ and its
exterior filled by $u_0\to-1$, which is exactly an $\pi_3[S^3]$
soliton of unit strength. The corresponding configuration of $U$
will be an SU(2) soliton of unit strength. The energy density
$(U^\dagger \nabla U)^2$ of the soliton is localized around the
origin, verifying that the soliton is an local excitation (in the
$\vec{u}$-ordered limit).

\begin{figure}[htbp]
\begin{center}
\includegraphics[width=180pt]{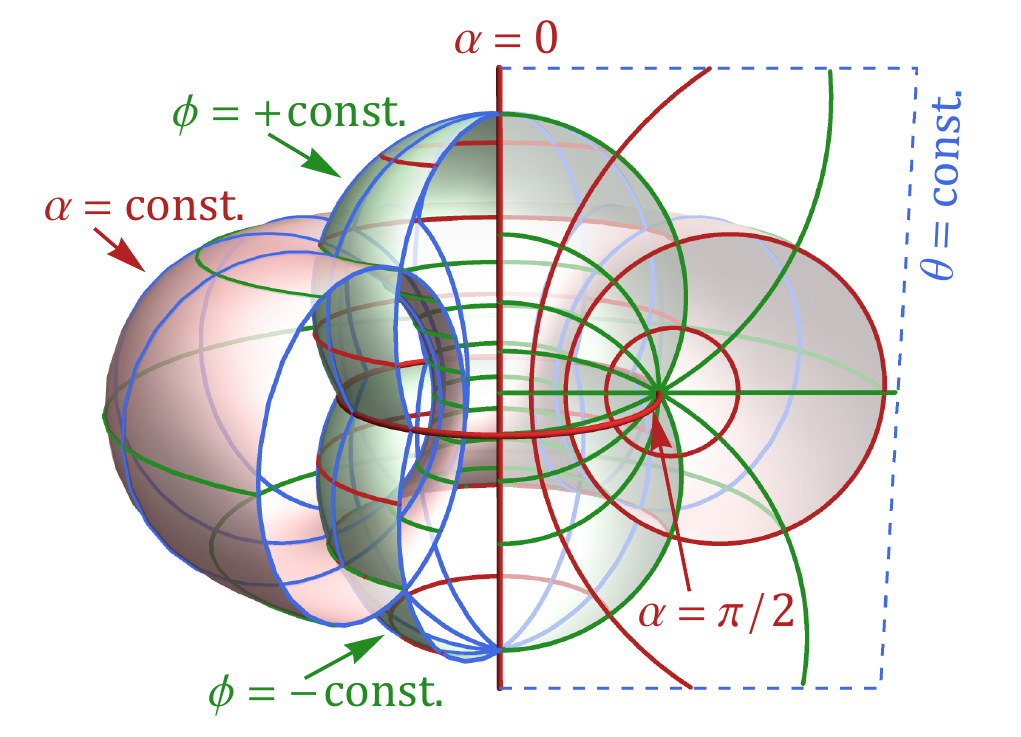}
\caption{Toroidal coordinate.}
\label{fig: toroidal}
\end{center}
\end{figure}

Consider an O(3) vector $\vect{n}=(n_1,n_2,n_3)\in S^2$ which
transforms as a spin-1 representation of the SU(2) group. Then the
SU(2) gauge transformation that creates an SU(2) soliton will
correspondingly create a Hopf soliton in the $\vect{n}$ field. To
see this, let us start from a trivial configuration of $\vect{n}$
with all the vectors polarized to $\vect{n}(\vect{r})=(0,0,1)$.
After the SU(2) gauge transformation induced by the field
$U(\vect{r})$, the configuration of $\vect{n}(\vect{r})$ will
become \beq\label{eq: n=TrUsUs} \vect{n}=\frac{1}{2}\Tr U^\dagger
\vect{\sigma} U \sigma^3. \eeq This is a Hopf map from the SU(2)
manifold to $S^2$, under which the SU(2) soliton is mapped to a
Hopf soliton. However it is difficult to visualize the Hopf
soliton in the Cartesian coordinate, thus we switch to the
toroidal coordinate $(\alpha,\phi,\theta)$, which is defined by
\beq\label{eq: toroidal coordinate}
\vect{r}=(x,y,z)=\frac{1}{\sec\alpha+\cos\phi}
(\tan\alpha\cos\theta,\tan\alpha\sin\theta,\sin\phi), \eeq where
$\alpha\in[0,\pi/2]$ and $\phi,\theta\in[-\pi,\pi)$ are the . The
geometric meaning of the toroidal coordinate is illustrated in
\figref{fig: toroidal}. In the new coordinate system, \eqnref{eq:
u(r)} is reduced to
$\vec{u}=(\cos\phi\cos\alpha,\cos\theta\sin\alpha,\sin\theta\sin\alpha,\sin\phi\cos\alpha)$.
Plugging into \eqnref{eq: U} and \eqnref{eq: n=TrUsUs} yields
\beq\label{eq: Hopf soliton} \vect{n}=\big(\sin(\phi-\theta)\sin
2\alpha, \cos(\phi-\theta)\sin 2\alpha, \cos 2\alpha\big). \eeq
The configuration can be described as follows: on the torus
specified by $\alpha=\pi/4$, $n_3=0$ and $(n_1,n_2)$ has a full
winding along both the meridian $\phi$ and the longitude $\theta$
directions; while the exterior (interior) of the torus is
gradually polarized to $n_3=+1$ ($n_3=-1$). In this configuration,
the preimages of $\vect{n}$ are mutually linked circles in the
$3d$ space, as shown in \figref{fig: Hopf}(b), so it is exactly a
Hopf soliton. Although the figure does not seem to be rotational
invariant but in fact the energy density $|\nabla\vect{n}|^2$ of
the Hopf soliton is spherical symmetric and localized around the
origin.

When the $\vect{n}$ field is coupled to the SU(2) Dirac fermion
$\psi$ in $3d$, \beq H=\int\dd^3 x\; \psi^\dagger
\ii\vect{\sigma}\cdot\vect{\partial} \psi+\vect{n}\cdot
\Re[\psi^\intercal \sigma^y\tau^y\vect{\tau}\psi], \eeq the Hopf
soliton will carries a fermion. More precisely, the creation of a
Hopf soliton will change the fermion parity locally. To see this,
we can first reduce the theory to the domain wall of $n_3$ on the
torus of $\alpha=\pi/4$, on which we further reduce the theory to
the domain wall of $n_1$ along the circle of $\phi-\theta=0$,
which reads $H=\int\dd
\xi\;\frac{1}{2}\chi^\intercal[\sigma^1(\ii\partial_\xi+\omega_\xi)+n_2\sigma^2]\chi$
in terms of the Majorana fermion $\chi$, where $\xi$ parameterize
the circle and $\omega_\xi$ is the spin connection along the
circle. It  turns out that the spin Berry phases along both the
meridian and the longitude directions are both $\pi$ on the
$\alpha=\pi/4$ torus, so the total Berry phase along the circle is
$2\pi$, meaning that the spin connection can be gauged away. Thus
the lowest momentum is quantized to $k=0$. On the circle, the Hopf
soliton is differed from a trivial configuration by the sign of
$n_2$: flipping $n_2$ from $n_2<0$  to $n_2>0$ corresponds to the
creation of the Hopf soliton, which, according to the effective
theory on the circle, will lead to a level crossing at $k=0$, and
hence change the fermion parity.

Due to the self-statistic of the fermion, the $2\pi$ rotation of
the Hopf soliton is expected to produce a minus sign in the
many-body wave function. In fact the Berry phase can be explicitly
calculated. To simplify, we can reduce the problem to the $n_3$
domain wall on the torus of $\alpha=\pi/4$, and then compute the
Berry phase accumulated over the $S$ modular transformation of the
torus, which is a $\pi/2$ rotation. It is found\cite{youcheng}
that the $S$ transformation will give a Berry phase of $\pi/4$, so
the full $2\pi$ rotation (four times of $S$ transformation) will
produce a minus sign in the wave function.

\subsection{B2. Vison loop and vison link}

In the $\dsZ_{2N}$ topological order phase, a vison line is a
$\pi$ gauge flux seen by the  CP$^1$ spinon
$z=(z_1,z_2)^\intercal$, meaning that the spinon going around the
vison line will acquire a minus sign as $z\to -z$. The vison line
will be bound with either a U(1) (gauge) half-vortex or an SU(2)
(symmetry) half-vortex. Assuming the vison line is along the axis
$\rho=0$ in a cylindrical coordinate $(\rho,\varphi,h)$, then the
U(1) and the SU(2) half-vortices are described respectively by
\beq \text{U(1): }z=e^{\ii\varphi/2}z_\text{ref},\quad
\text{SU(2): }z=e^{\ii\varphi\sigma^3/2}z_\text{ref}, \eeq where
$\varphi\in[0,2\pi)$ is the azimuthal angle around the vison line
and $z_\text{ref}=(z_1,z_2)_\text{ref}^\intercal$ is a spinon
reference state. As $\varphi$ goes from $0$ to $2\pi$, $z$ will
get a minus sign under both vortex configurations. However in the
$\dsZ_{2N}$ topological order phase, the U(1) gauge vortex is
gapped by the Higgs mechanism (because it corresponds to a vortex
in the $z_1z_2$ field, which is condensed in the $Z_2$ topological
order phase). So the SU(2) vortex is energetically favored around
the vison line. In the following, we will focus on the case that
the vison line is always bound with the SU(2) vortex.

To investigate the vortex link, it will be convenient to switch to
the toroidal coordinate defined in \eqnref{eq: toroidal
coordinate}. Let the two vison loops (or lines) be the vertical
axis $\alpha=0$ and the horizontal ring $\alpha=\pi/2$ in the
toroidal coordinate. The link of the SU(2) half-vortices of $z$
can be described by \beq z=e^{\ii(\theta-\phi)\sigma^3/2}e^{-\ii
\alpha\sigma^1} z_\text{ref}. \eeq The operator
$e^{\ii(\theta-\phi)\sigma^3/2}$ impose the SU(2) rotation by
$\pi$ in both the meridian and the longitude directions. As either
$\theta$ or $\phi$ going from $0$ to $2\pi$, the spinon $z$ will
get a minus sign as required by the vison loops. Then in terms of
the order parameter $\vect{n}=z^\dagger\vect{\sigma} z$, the
configuration will be a pair of SU(2) vortices linked together
\beq \vect{n}=e^{\ii(\theta-\phi)J_3} e^{-2\ii\alpha J_1}
\vect{n}_\text{ref}, \eeq where $(J_i)_{jk}=\ii\epsilon_{ijk}$
($i,j,k=1,2,3$) are the generators of SO(3). To avoid singularity
in the configuration of $\vect{n}$ (otherwise there will be
gapless fermion modes), we must have
$\vect{n}_\text{ref}=(0,0,\pm1)$. Suppose we choose
$z_\text{ref}=(1,0)^\intercal$ and $\vect{n}_\text{ref}=(0,0,1)$,
then
$\vect{n}=\big(\sin(\phi-\theta)\sin2\alpha,\cos(\phi-\theta)\sin2\alpha,\cos2\alpha\big)$
will exactly be the Hopf soliton configuration given in
\eqnref{eq: Hopf soliton}. Thus we conclude that the SU(2) vortex
link (bound to the vison link) is equivalent to a Hopf soliton,
which, after coupling the  order parameters to the fermions, will
also carry a fermion as required by the Witten anomaly.

\end{document}